\documentclass{elsart}

\usepackage{amssymb}
\usepackage{epsf,colordvi,color,amsbsy}
\usepackage[dvips]{graphicx}
\begin{document}

\begin{frontmatter}

\title{Solar cycle full-shape predictions: a global error evaluation for cycle 24}

\author[]{Stefano Sello \corauthref{}}

\corauth[]{stefano.sello@enel.it}

\address{Mathematical and Physical Models, Enel Research, Pisa - Italy}

\begin{abstract}

There are many proposed prediction methods for solar cycles behavior. In a previous paper we updated the full-shape curve prediction of the current solar cycle 24 using a non-linear dynamics method and we compared the results with the predictions collected by the NOAA/SEC prediction panel, using observed data up to October 2010.
The aim of the present paper is to give a quantitative evaluation, a posteriori, of the performances of these prediction methods using a specific global error, updated on a monthly basis, which is a measure of the global performance  on the predicted shape (both amplitude and phase) of the solar cycle. We suggest also the use of a percent cycle similarity degree, to better evaluate the predicted shape of the solar cycle curve.

\end{abstract}
\end{frontmatter}

\section{Introduction}

Solar cycle full-shape prediction methods aim to determine the approximate whole cycle curve, not only its peak magnitude and timing. This task is particularly useful for numerous
scientific and technological applications. The main involved areas are the electric power transmission systems, airline and satellite communications, GPS signals, and extra-vehicular-activities of astronauts during space missions and, more in general, all the solar-terrestrial interactions.
Numerous techniques for time series forecasting are developed to accurately predict phase and amplitude of future solar cycles, but with limited success. Depending on the nature of the prediction methods we can distinguish five principal classes: 1) Curve fitting; 2) Precursor; 3) Spectral; 4) Neural Networks; 5) Climatology.
Apart from precursor methods, the main limitation is
the short time interval of reliable extrapolations, as the
case of the McNish-Lincoln curve fitting method. In
the climatological method we predict the behaviour of the
future cycle by a weighted average from the past N cycles,
based on the assumption of a some degree of correlation
of the phenomenon. 
Modern classes of solar activity prediction methods appear
to be reasonably successful in predicting `long range'
behavior, the so-called precursor methods. Precursors
are early signs of the magnitude of future solar activity
that manifest before clear evidence of the next solar cycle.
There are two main kinds of precursor methods: geomagnetic and solar. The basic idea is that if these methods work,
they must be based on solar physics, in particular dynamo
theory. The precursor methods are correlated with a solar
dynamo mechanism, where we suppose that the polar field
in the descending phase and minimum of a cycle gives an
indication of future fully developed toroidal fields inside
the Sun that will drive the next solar activity. Although the dynamo methods and, in general, the precursor methods seems to work quite well, they might be affected
by some severe drawbacks, mainly depending on various unknown physical parameters. However, as pointed out
by different authors, we need again a better scientific basis. Thus, at present, the statistical-numerical approaches, based on reliable characterization and prediction of complex time series behavior, without any intermediate physical model, still appear as valuable techniques to provide at least the basic elements for new insight for future advanced physical prediction methods. Many different techniques proposed in literature appear quite reliable and accurate to predict, well in advance, the peak and timing values of future solar cycles.
On the other hand, it is well known the difficulties in predicting the full evolution of future solar cycles, due to highly complex non-linear dynamical processes involved, mainly related to interaction of different components, deterministic and stochastic ones, of the internal magnetic fields.
We recall here some brief introductive considerations of the argument under study, as well reported in the reference article by Sello, 2012a.  
The predictions of the "anomalous" current solar cycle 24 allow us to deeply test our current knowledge on the complex dynamics of solar activity evolution and to further refine both our current solar dynamo models and new mathematical/numerical methods.
The main detected anomaly of cycle 24 is related to its extended period of minimal solar activity, with very few small and weak sunspots and flares. There is now a general consensus about a clear diminished level of solar activity, suggesting a current transition phase between a long period of maximum and a next long period of moderately or weak activity (see: de Jager and Duhau, 2009). In the modern era there is no precedent for such a protracted activity minimum, but there are historical records from a century ago of a similar pattern. 
Based on the flux F10.7 activity index, there seems to be little question that the new-cycle activity 24, started late in 2008 or in early 2009. 
The successive progression of the solar cycle was very slow but constant and a first peak of strong activity is well documented during year 2011, with the detection of strong X-flares activity and large sunspot areas
with complex magnetic field configuration (see Fig.1 showing the big region AR11302).

\begin{figure}[!ht]
\centering
\includegraphics[scale=1]{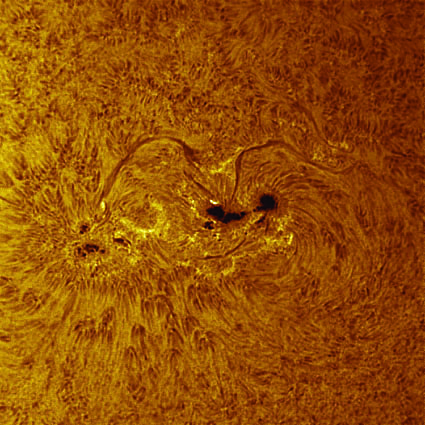}
 \caption{Big region $AR11302$ taken through a line centered $H\alpha$ filter. Beta-gamma-delta magnetic configuration and area: $750$ msh and located at $N13W16$ (H.A.S.O. September 29, 2011 15:30 UT).}
 \label{fig1}
\end{figure}

The unusual character of the current solar cycle is also evidenced by the unexpected results of the prediction methods documented in the literature. A comprehensive collection of predictions for solar cycle 24 is described by Pesnell and regularly released by NOAA/SEC international panel (Pesnell, NOAA/SEC panel).
The considered prediction methods include different classes: Precursors, Spectral, Climatology, Neural Network, Physics based, Statistical, etc.
In the panel forecast for sunspot numbers released in 2009, half of the members predicts a moderately strong cycle of $140 \pm 20$ expected to peak in October 2011; the other half predicts a moderately weak cycle of $90 \pm 10$ peaking in August 2012.
If we consider a small but representative sample of different forecasts we clearly obtain the uncertainty of the whole set of predictions included in the NOAA/SEC panel document and explains the evenly split between a moderately strong and a moderately weak solar cycle.
Moreover, as the cycle progresses there is an increasing consensus towards a weak solar activity.
For more details on the prediction results from different methods, including the nonlinear one, we refer to paper by Sello, 2012a.
In order to quantitatively evaluate the performances of forecasts for solar cycle 24, we need to compute a global error index, updated on a monthly basis, which gives a performance evaluation on the whole shape predicted curve as the cycle progresses. Unfortunately, it is not very common to find works in the literature specifically devoted to verify the performances a posteriori of previously proposed forecasting methods, also because this process requires a long term analysis using the observed data for a period of, at least, five or six years after the given prediction. In fact, many authors, after the publication of their predictions during the increasing phase of the solar cycle, forget to evaluate a posteriori, several years later, the quality of the suggested methods.
Here we compared the nonlinear method prediction with the NOAA/SEC panel prediction which is the better estimate coming from many reliable and accurate classes of prediction methods. A preliminary list of predictions from different methods can be found in a report by Pesnell, NASA, Goddard Space Flight Center, May 2007. More updated reviews have been published in 2008 and 2012.
Since Waldmeier work (1935), several authors have proposed to represent the shape of the solar cycle using families of analytical models with few free parameters, such as the start time and amplitude of the cycle, in order to estimate the mean trend of both the ascending and descending phases of the cycle. These parametric fully deterministic analytical models have, as main objective, to provide a mean behavior of the cycle or a general curve shape fitting. We note that some of the free parameters must be carefully tuned in these models and it is not always easy to accurately estimate their values near the minima. However, here we consider the result shown by Wilson, 2011. Using the 'HWR' two free parameters Planck-like general shape fitting function with the observed first 18-months sunspot numbers (12-month moving average) of cycle 24 (end 2010), the author obtains a best-fit prediction for monthly smoothed sunspot numbers peak of 70 at August 2013. This fairly good result is consistent with the "low" panel consensus prediction (Pesnell, 2008).

\section{Global error evaluation}

The global error evaluation for the whole shape (amplitude and phase) of the solar cycle was already suggested in Sello, 2001 for previous solar cycle 23.
Here we recall shorthly the main definition only. The accuracy of a full-shape prediction model is estimated through the computation of the "global average prediction error" defined as:

\begin{equation}
    <E^2>(t)=1/{n_t} {\sum_{i=1}^{n_t} (y_i^{pr}-y_i^{obs})^2 / {\sigma^2}}
\end{equation}

where $y_i^{pr}$ and $y_i^{obs}$ are the predicted and observed monthly smoothed sunspot numbers, respectively and $n_t$ is the number of observations related to time t. Here $\sigma$ is the standard deviation of observed data. 
The global error can be updated on a monthly basis in order to see the behavior of the error as the cycle progresses. Of course, the optimal model corresponds to a minimum value of
$<E^2>(t)$ as a function of t. However, the final comparison of different prediction methods is given through the global error evaluation for the last t value which contains a more comprehensive cycle information.
For further details the reader may refer to: Sello, 2001.

\section{Results}

Figure 2 shows the monthly smoothed predictions released in the original article (see: Sello, 2012a), comparing nonlinear method with NOAA/SEC prediction, using observed data from SIDC, RWC Belgium up to October 2010, when the initial increasing phase of activity is already well developed. Of course, successive updated and revised predictions are certainly more accurate but here we are interested to show the full forecast potential of a given full-shape prediction method, where we are again well distant from the next maximum phase. However, in order to make a correct evaluation of different forecasting methods it is necessary to take into account only predictions made in the same period, with the same available information. 

\begin{figure}[!ht]
\centering
\includegraphics{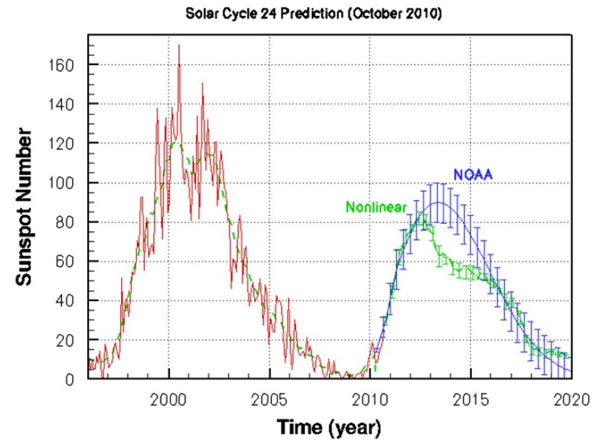}
 \caption{The full-shape monthly smoothed predictions from nonlinear method (green solid line) and NOAA/SEC panel (blue solid line) with error bars using observed data from SIDC up to October 2010 (red line).}
 \label{fig2}
\end{figure}

At the beginning we see a clear delay in the early ascent period of the cycle than predicted, determining a significant initial error of predictions both on amplitude and phase.

In the following Figure 3 we show the results of the global error computation as a function of time, on a monthly basis, as the solar cycle progresses. Here we can directly compare the global error with the corresponding solar cycle phase.
The observed monthly smoothed sunspot numbers are provided by WDC-SILSO, Royal Observatory of Belgium, Brussels. As stated in the related web site, on July 1st, 2015, the sunspot number series has been replaced by a new improved version (version Nb. 2.0) that includes several corrections of past inhomogeneities in the time series. The most prominent change in the Sunspot Number values is the choice of a new reference observer, A.Wolfer (pilot observer from 1876 to 1928) instead of R. Wolf himself. This means dropping the conventional 0.6 Zurich scale factor, thus raising the scale of the entire Sunspot Number time series to the level of modern sunspot counts. More information about the various diagnostics and corrections applied to the Sunspot Number and Group Number series in version 2.0 can be found in: Clette et al.(2014). 
The residual differences from the two time series are mainly located after 1950 with an irregular pattern that mimics the solar cycle behavior. In order to consistently perform our analysis even past July 1st 2015, we reconstructed the original older values using a combined method, observed/predicted, that uses both the new observed values and a predicted set of differences, through a properly designed Neural Network predictor. This allowed us to use, as a reference and without interruption, a consistent monthly smoothed sunspot number time series up to January 2016 (available in June 2016), where the global average prediction error value is well stabilized.

\begin{figure}[!ht]
\centering
\includegraphics{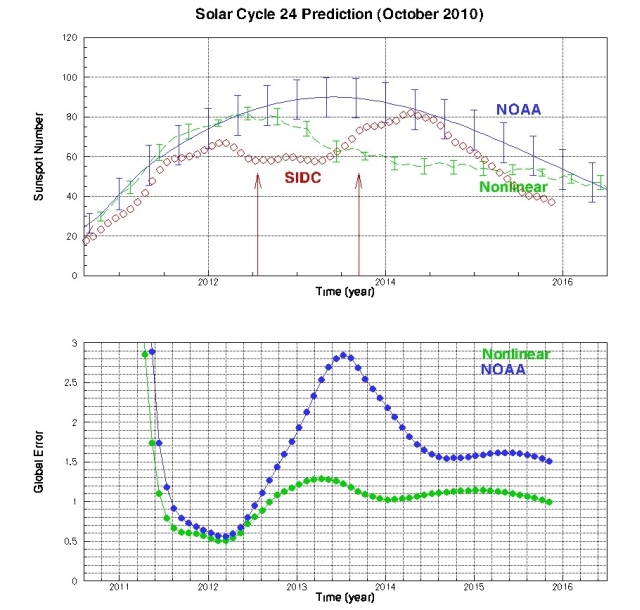}
 \caption{Predicted and observed (SIDC data) monthly smoothed sunspot numbers (up) and the global error behavior for nonlinear prediction method (green) and for NOAA/SEC prediction (blue) computed on a monthly basis (down). See also Fig.2. The two red arrows indicate the time of poloidal field reversals: first north and south after. The observed monthly smoothed values are provided by WDC-SILSO, Royal Observatory of Belgium, Brussels.}
 \label{fig3}
\end{figure}

A further unexpected delay in the first increasing phase of the cycle has most affected both the prediction methods; however, the subsequent recovery is quite rapid. The nonlinear method starts better with values less distant from the observed data. However, as the cycle progresses the two methods tend to reach an equivalent error also due to some overestimated increasing values from nonlinear method. It is already noted that there are clear evidences that the solar maximum, given by the monthly smoothed sunspot numbers, in the northern hemisphere occurred at $2011.6 \pm 0.3$ (with the value: $41.17$), as we can see from the quite constant behavior of the observed data (see also: Sello, 2012b). After a short increasing phase, the observed data start to decrease since 2012.3 suggesting a monthly smoothed maximum reached at 2012.2 with the value: $66.9$ (monthly mean maximum reached at 2011.9 with the value: $96.7$), well in advance with respect to predictions. This caused an increase of the updated global error for both the methods. We recall here that the nonlinear method predicted a monthly smoothed maximum at around 2012.5 with the value: $82\pm 5$; whereas the NOAA/SEC predicted a maximum value of $90\pm 10$ at around 2013.4. After 2012.45 the two prediction methods start to deviate: in particular the nonlinear method starts the decreasing phase of the monthly smoothed values, whereas the NOAA prediction confirms a stable increasing phase of the amplitude values. This fact leads to significant effects on the global average prediction error. The increasing rate of the first derivative is quite clear from the figure. In particular, after 2012.83 an inflection point reduces the increase of nonlinear global error curve, whereas the NOAA global error curve maintains a nearly constant value of its first derivative. A second smaller peak for monthly smoothed values is visibile near 2012.9 with a value around $60$. It is well known the multi-peaked structure of many solar cycle maxima and the related Gnevyshev gap (see: Sello, 2003). A new strong increasing trend is visible after 2013.38 towards a greater monthly smoothed maximum reached at 2014.3 with the value: 81.9 (monthly mean maximum reached at 2014.1 with the value: 102.3). This is the updated main peak of solar cycle 24. The multipeaked structure of the maximum phase of solar cycle 24 is now clearly visible. This feature is very difficult to predict in detail using the current prediction  methods. In fact, after 2014.1 the global error of nonlinear method starts to increase whereas the global error of NOAA/SEC quickly reduces to smaller values.
The predicted amplitude of the monthly smoothed maximum is quite good for nonlinear method: $82\pm 5$ vs $81.9$  (only $0.12\%$ higher) but unfortunately the predicted phase is quite wrong (see figure 3). One possible explanation can be the times of reversals of the polar poloidal field that occur near the maxima of sunspot activity. In fact, the current solar cycle 24 is characterized by a relatively high north-south asymmetry of
polar field reversal: the northern polar field reversed its sign more than one year before the southern field (13.68 months). It is well known that fluctuations are inherent to the Babcock-Leighton $\alpha$-effect and the times of polar field reversals on both hemispheres are quite difficult to predict (see: Kitchatinov and Khlystova, 2014  and references therein). A high north-south asymmetry in the reversal of poloidal fields produces a delay in the evolution of solar maximum phase, with a longer duration of the Gnevyshev gap. The consequence is an unpredictable delay in the developing of the maximum phase peaks.
Thus solar cycle 24 proves to be $32\%$ weaker than the previous solar cycle and thus belongs to the class of moderate cycles, like cycles 12 to 16, which were usual in the late 19th and early 20th century. In fact, compared to recent strong cycles, such moderate cycles typically feature a broader maximum, with a multi year plateau on top of which two or more surges of activity can produce multiple peaks of similar height.
As a comparison, for the previous solar cycle 23, the nonlinear method predicted a monthly smoothed peak of $3.9\%$ higher than the observed value, well inside the predicted uncertainty (Sello, 2001). 

The last value of the global average prediction error computed up to: $T=2015.85$ (i.e. excluding the common first rising and the last descending phases of the cycle, outside the so-called extended maximum phase of a solar cycle) is: $0.99$ and $1.5$ for nonlinear and NOAA/SEC methods, respectively. These values quantify the global prediction error on the whole curves in the considered time interval (about $5$ years). The normalized mean values related to a unit time interval (i.e. the slopes of the equivalent linear behaviors) are: 0.19 $yr^{-1 }$ and: 0.3 $yr^{-1 }$ for nonlinear and NOAA/SEC respectively.
    
These values suggest that solar cycle $24$ was a quite difficult task for solar cycle predictors. As a comparison, for previous solar cycle 23, the global average prediction error of nonlinear method reached a normalized mean value of 0.14 $yr^{-1 }$ (Sello, 2001). 

To complete the solar cycle full-shape prediction evaluation, we add to the global average prediction error, eq.(1), a curve similarity index derived from spectroscopy comparison methods.
More precisely, taking a lead from the spectral contrast angle methods used in HPLC-UV spectroscopy, Wan et al. 2001, implemented vector based representations of tandem mass spectra. A curve or spectrum is a N-dimensional vector constructed with N different predicted or observed values at the same time or frequency values. Here we use the Euclidean distance to evaluate the lenght of vectors. The lengths (r) of two vectors A and B  $\in {R^N} $ are simply determined by:

\begin{equation}
    r_a= \sqrt{\sum_{i=1}^{N} a_i^2}
\end{equation}

\begin{equation}
    r_b= \sqrt{\sum_{i=1}^{N} b_i^2}
\end{equation}

The two curves (vectors) can be quantitatively compared by the
derived spectral contrast angle ($\theta$). The angle, $\theta$, is defined as:

\begin{equation}
    cos(\theta)= {\sum_{i=1}^{N} a_i b_i \over \sqrt{ {\sum_{i=1}^{N} a_i^2 }  {\sum_{i=1}^{N} b_i^2 } } }
\end{equation}

where $a_i$ and $b_i$ are the intensities or amplitudes of the corresponding curves A and B. An angle of zero degrees means there are no discernible curve differences. Curves that resemble each other have vectors that point in the same direction in the space. A $90$ degree angle indicates a maximal curve differentiation. The similarity index is a real absolute number between 0 (maximum difference) and 1 (maximum similarity) and it can be expressend as a percent similarity.
Here we computed a percent cycle similarity degree index based on the above spectral contrast angle comparing the shape of a given prediction curve (A) with the corresponding curve shape of observed values (B). Considering the current whole set of $N=62$ data in Fig.3, the percent cycle similarity degree (PCSD) is: $0.01\%$ and $0.19\%$ for nonlinear and NOAA/SEC prediction curves, respectively. The very low value of similarity index obtained for nonlinear method is mainly due to the presence of the second and dominant peak in the solar cycle activity, producing an opposite phase behavior in the observed and predicted curves. Another interesting full-shape cycle prediction method, based on a similar cycle concept developed by Du and Wang, 2011 (using both the rising rate and solar minimum value to select the similar cycles), predicted monthly smoothed sunspot numbers for cycle 24, using SIDC data up to November 2010, obtaining a peak around January $2013 \pm 8$ (months) with a size of about $Rmax = 84 \pm 17$. If we compute the percent cycle similarity degree for the whole curve (see Fig. 5c in the cited paper), we found the value: $0.12\%$. On the other hand, the cited 'HWR' general shape fitting method (Wilson, 2011), gives a  percent cycle similarity degree for the whole curve of $4.37\%$, a fairly good relative result. In order to better evaluate the shape similarity of the different prediction methods, we computed the time behavior of index PCSD, starting from 12 months after the epoch: October 2010 (2010.788). Figure 4 shows the PCSD time evolutions for the above four shape prediction methods: nonlinear, NOAA, Wang and HWR.

\begin{figure}[!ht]
\centering
\includegraphics{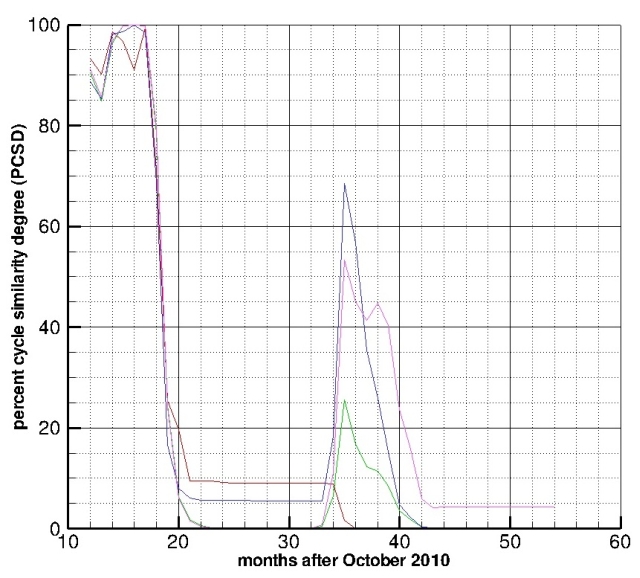}
 \caption{The PCSD time evolution for four prediction methods considered in this paper. red: nonlinear; green: NOAA; blue: Wang, magenta: HWR}
 \label{fig4}
\end{figure} 

After an initial high PCSD value for all methods due to the first initial phase of the cycle, there is a sharp decreasing trend of the similarity index for all methods up to 20-months after the selected epoch. In the interval 20-33 months after the epoch the nonlinear method shows the better performance, with a PCSD value around 10\%. But after 34-months, the similarity index behavior completely changes: we note a sharp decrease of PCSD for nonlinear method and a fast increasing trend for NOAA, HWR and Wang predictions with peaks: 25\%, 58\% and 68\%, respectively. This phase corresponds to the second increase trend of the solar activity towards the second and higher peak of the cycle. After the local peak the behavior of  PCSD index changes again with a new fast and deep decreasing trend for all methods and after about 54-months from the epoch only the HWR prediction reaches an acceptable far from zero stable similarity degree index, around 4\%.  

Although the HWR method has performed better in predicting the global shape of the solar cycle curve than the other methods considered here, the low value of the PCSD found ($4.37\%$), suggests a quite different full shape, as in fact we can observe, and that we are still too far from having a method for predicting the full cycle curve that is sufficiently accurate. It will need further deep work on the basis of both precursors and numerical-statistical methods, as well as an upgrade of the current physical models based on both solar dynamo and stochastic mechanisms (e.g. fluctuations of the meridional flow, see: Sch\"{u}ssler, 2007) in order to obtain quite reliable and accurate prediction methods not only for the amplitude and phase of the maximum peak but also for the solar cycle full-shape curve. The degree of predictability of solar cycles is still a great subject of an extensive discussion inside the solar physics community and it is not yet clear whether it will be possible to model the overall multi-scale components involved in the physical processes, both deterministic and stochastic, responsible for the cyclical solar activity.  (see: Bushby and Tobias, 2007; Abreu et al., 2008; for a recent review see: Balogh et al., 2014). It is quite significant a recent work by Nandy and Karak, 2013, where, considering the turbulent pumping as the dominant mechanism for flux transport, it is possible to show that the solar cycle memory (here defined as the finite time necessary for the coupling of the spatially segregated source layers of the dynamo mechanism) is short, lasting less than a complete 11 year cycle and this implies that solar cycle predictions for the maxima of cycles are best achieved near the preceding solar minimum, about 4-5 years in advance, whereas  long-term predictions are unlikely to be accurate. This result is quite consistent with previous estimations of the predictability horizon (tp = 4.85 years) based on nonlinear dynamics concepts (Sello, 2001).

\section{Conclusions}

A method of performance evaluation of a given full-shape solar cycle curve prediction is proposed, computing a running global average prediction error and a specific percent cycle similarity degree. The unusual character of the current solar cycle 24 has put a strain on even the best consolidated forecasting methods. Here we compared the precursor non-linear method and the NOAA/SEC panel prediction using monthly sunspot number observed data up to October 2010. The evaluation of the full-shape cycle predictions is here performed considering the time interval: 2010.8-2015.85 The performance evaluation can be summarized by two indices: the last value of the global average prediction error computed on the whole curve and the corresponding percent cycle similarity degree. The first index was: 0.99 and 1.5 for nonlinear and NOAA/SEC prediction methods respectively. The second index was: 0.01 and 0.19 for nonlinear and NOAA/SEC prediction methods respectively. The best relative performance has been reached by HWR method with a similarity degree index of about 4. Although none of the considered methods produced a sufficiently high value of the similarity degree, the small amplitude of the monthly smoothed peak is quite well predicted by various prediction methods, but due to the strong multipeak shape, with a significant wrong phase. In fact, this is the main reason for a very low value of the similarity index obtained using the considered methods. The second strong peak in the solar cycle curve is the main factor for a very low value of similarity index for the nonlinear method. Further deep work is needed in order to obtain quite reliable and accurate prediction methods of the solar cycle full-shape curve, both amplitudes and phases: for this task we suggest the use of the above two performance indices, the progressive values of the global average prediction error computed on the whole curve and the corresponding percent cycle similarity degree, to better evaluate and to quantitatively compare the "a posteriori" performances obtained from different methods developed for future solar cycles predictions.

\section{References}

Abreu, J. A., J. Beer, F. Steinhilber, S. M. Tobias, and N. O. Weiss, 2008:  For how long will the current grand maximum of solar activity persist?, Geophys. Res. Lett., 35, L20109.

 Balogh, A.,  Hudson, H.S., Petrovay, K., von Steiger, R., 2014: Introduction to the Solar Activity Cycle: Overview of Causes and Consequences, Space Sci Rev, 186:1–15

Bushby, P. J., and Tobias S. M., 2007: On predicting the solar cycle using mean-field models, Astrophys. J., 661, 1289– 1296.

Clette, F., Svalgaard, L., Vaquero, J.M., Cliver, E. W., 2014: Revisiting the Sunspot Number. A 400-Year Perspective on the Solar Cycle, Space Science Reviews, Volume 186, Issue 1-4, pp. 35-103.

de Jager, C. and Duhau, S., 2009: Forecasting the parameters of sunspot cycle 24 and beyond, Jour. Atm. Solar-Terr. Phys., 71, 239.

Du, Z.L. and Wang, H.N. 2011, The prediction method of similar cycles, arXiv:1110.6436v1[astro-ph.SR].

Kitchatinov, L.L., Khlystova, A.I., 2014: North-south asymmetry of solar dynamo in the current activity cycle, arXiv:1406.7072v1 [astro-ph.SR]

Nandy, D., Karak, B.B., 2013: Forecasting the solar activity cycle: new insights, Solar and Astrophysical Dynamos and Magnetic Activity
Proceedings IAU Symposium No. 294.

NOAA, Space Weather Prediction Center: http://www.swpc.noaa.gov/SolarCycle/

Pesnell, W.D., 2007: Predictions of solar cycle 24, Solar Cycle 24, NOAA/SEC panel report, May 24, 2007; Predictions of Solar Cycle 24, 2008, Solar Phys, 252: 209-220; Solar cycle predictions. Sol. Phys. 281, 507–532 (2012).

Sch\"{u}ssler, M., 2007: Are solar cycles predictable?, Astron. Nachr. / AN 328, No. 10, 1087 – 1091

Sello, S., 2001: Solar cycle forecasting: A nonlinear dynamics approach, Astronomy and Astrophysics, 377, 1, 312.

Sello, S., 2003: Wavelet entropy and the multi-peaked structure of solar cycle maximum, New Astronomy, Volume 8, Issue 2, p. 105-117

Sello, S., 2012a: A nonlinear full-shape curve prediction after the onset of the new solar cycle 24, Jour. of Atmos. and Solar-Terrestrial Physics, 80 252-257.

Sello, S., 2012b: Solar Cycle 24: is the peak coming?, arXiv:1209.6135v2 [astro-ph.SR].

SIDC-team, World Data Center for the Sunspot Index, Royal Observatory of Belgium, Monthly Report on the International Sunspot Number, online catalogue of the sunspot index: http://www.sidc.be/sunspot-data/,�year(s)-of-data�.

Waldmeier , M. : 1935, Astr. Mitt. Zurich, 14, 133, 105.

Wan, K.X., Vidavsky, I. and Gross, M.L. 2001: Comparing Similar Spectra: From Similarity Index to Spectral Contrast Angle, American Society for Mass Spectrometry.

Wilson, R.M., 2011: An estimate of the size and shape of sunspot cycle 24 based on its early cycle behavior using the Hathaway-Wilson-Reichmann shape-fitting function, NASA Technical Paper, 216461, Marshall Space Flight Center, AL, USA, March, 2011.

\end{document}